\documentclass[reprint,aps,prl,twocolumn,superscriptaddress,amsmath,amssymb,longbibliography]{revtex4-2}
\usepackage{graphicx}
\usepackage[colorlinks=true,citecolor=blue,
            linkcolor=blue,urlcolor=blue,bookmarks=true]{hyperref}
\usepackage[utf8]{inputenc}
\usepackage[english]{babel}
\usepackage{amsmath}
\usepackage{xcolor}
\usepackage{soul}
\usepackage{mathbbol}
\usepackage{multirow}
\usepackage{physics}
\usepackage{algpseudocode}
\newcounter{algorithm}
\renewcommand{\thealgorithm}{\arabic{algorithm}}

\makeatletter
\renewcommand{\fnum@figure}[1]{\textbf{FIG.~\thefigure.~}}
\makeatother

\definecolor{norange}{RGB}{230,120,20}

\newcommand{\medTTe}{\left[\mathrm{TT}\varepsilon\right]_{\mathrm{Med}}}
\newcommand{\TTe}{\mathrm{TT}\varepsilon}
 
\begin{document}
\raggedbottom 
\title{Toward quantum scaling advantage in approximate optimization}

\author{J. Paw\l{o}wski}
\affiliation{Institute of Theoretical Physics, Faculty of Fundamental Problems of Technology, Wroc\l{a}w University of Science and Technology, 50-370 Wroc\l{a}w, Poland}

\author{P. Tarasiuk}
\affiliation{Quantumz.io Sp.\;z\;o.o., Puławska 12/3, 02-566 Warsaw, Poland}

\author{J. Tuziemski}
\affiliation{Quantumz.io Sp.\;z\;o.o., Puławska 12/3, 02-566 Warsaw, Poland}
\affiliation{Institute of Informatics, Faculty of Mathematics, Physics and Informatics, University of Gdańsk, Wita Stwosza 57, 80-308 Gdańsk, Poland}

\author{Ł. Pawela}
\affiliation{Institute of Theoretical and Applied Informatics, Polish Academy of Sciences, Ba{\l}tycka 5, 44-100 Gliwice, Poland}
\affiliation{Quantumz.io Sp.\;z\;o.o., Puławska 12/3, 02-566 Warsaw, Poland}

\author{B. Gardas}
\affiliation{Institute of Theoretical and Applied Informatics, Polish Academy of Sciences, Ba{\l}tycka 5, 44-100 Gliwice, Poland}

\begin{abstract}
In a recent Letter~\cite{Lidar2025}, quantum annealing was reported to exhibit a scaling advantage in approximately solving
quadratic unconstrained binary optimization (QUBO) problems. Here, we revisit these findings by employing the simulated bifurcation
machine (SBM), a nonlinear dynamical system that exploits chaotic behavior rather than thermal fluctuations. Our approach originates from quantum
dynamics and shares key operational features with quantum annealing: (i) nearly parallel evolution and (ii) a well-defined
relation between the energy gap, run-time, and solution quality. We obtain comparable or superior scaling, closing
the reported quantum--classical gap. We further show that the small instances studied previously are insufficient to infer asymptotic behavior. 
Extending the analysis to larger problems reveals robust classical performance, indicating that current
quantum annealers are unlikely to exhibit a clear scaling advantage over SBM-like solvers on quantum-annealing-correction-type QUBO problems under the run-time accounting studied here.
Finally, we identify sparse problem classes where future quantum devices could achieve a genuine scaling advantage,
once hardware overheads are mitigated.
\end{abstract}

\maketitle

{\bf Introduction --}
A central goal of quantum computing is to demonstrate quantum advantage -- tasks where quantum devices 
outperform classical ones~\cite{advantageReview,advantageCharacterization}. 
Despite major progress, a clear demonstration remains elusive~\cite{RCSGoogle2024}. 
Quantum advantage may manifest as reduced complexity, faster run-time, or lower energy use, 
but any claim must be validated against steadily improving classical algorithms~\cite{QuantumComputerSimulations,Spoofingadvantage,SpoofingadvantageSTOC,IsingIBMSimulation}.
This was evident in random circuit sampling~\cite{Sycamore}, where an initial gap was soon narrowed~\cite{IBMSycamore} 
and later closed~\cite{SycamoreSolution} by classical methods. 
For current NISQ devices~\cite{Preskill2018quantumcomputingin}, such claims tend to provoke debate rather than consensus. 
A similar pattern followed a recent ``quantum advantage'' claim in physical simulation~\cite{DWaveadvantage}, 
which prompted rapid advances in tensor-network approaches~\cite{DWaveadvantageComment1,DWaveadvantageComment2} 
and continued scrutiny~\cite{DWaveadvantageResponse}.

A more recent claim~\cite{Lidar2025} reported a quantum advantage in approximate optimization, 
where D-Wave quantum annealing was said to outperform the ``\emph{top classical heuristic algorithm}'' on Quadratic Unconstrained 
Binary Optimization (QUBO) problems~\cite{Glover2022}. QUBO is equivalent to the Ising model, 
\(H = -\sum_{i<j} J_{ij}s_is_j - \sum_i h_i s_i\) with binary spin variables $s_i = \pm 1$, couplings $J_{ij}$, and local fields $h_i$. 
Thus, such problems are naturally suited for quantum annealers~\cite{zephyr}. Given their broad practical relevance~\cite{QUBOSurvey}, the 
claim deserves careful reexamination.

In this Letter, we explain why the reported advantage cannot hold and provide a
conceptual foundation for advancing toward genuine quantum advantage in
approximate optimization. We examine whether a classical algorithm sharing key
features of quantum annealing, namely nearly parallel dynamics and a
well-defined relation between the energy gap, run-time, and solution
quality, can reproduce similar scaling. The answer is affirmative: the favorable
scaling reported in Ref.~\cite{Lidar2025} cannot be uniquely interpreted as 
quantum advantage. Instead of thermally or stochastically inspired solvers such
as PT-ICM, we employ a classical method obtained via a dequantization
procedure, the simulated bifurcation machine (SBM)~\cite{Goto2016}. This approach
is based on nonlinear Hamiltonian dynamics and provides a quantum-inspired
classical baseline.

\emph{Our approach yields scaling comparable to, and in some regimes better than,
that of the quantum annealer, removing the previously reported gap to a classical baseline.}
We further highlight methodological assumptions and finite-size effects in Ref.~\cite{Lidar2025} that biased results toward quantum annealing. 
Extending the analysis to much larger problems, with up to $N \approx 4 \times 10^4$ variables
(requiring a QPU with Pegasus topology and at least $1.5 \times 10^5$ physical qubits), 
we establish a robust classical baseline for future quantum advantage studies. 
Finally, we identify sparse instances where quantum annealers can rapidly obtain high-quality solutions and may demonstrate 
a genuine scaling advantage \emph{once hardware-level run-time limitations are mitigated.}

\begin{figure*}[htbp]
    \centering
    \includegraphics[width=\textwidth]{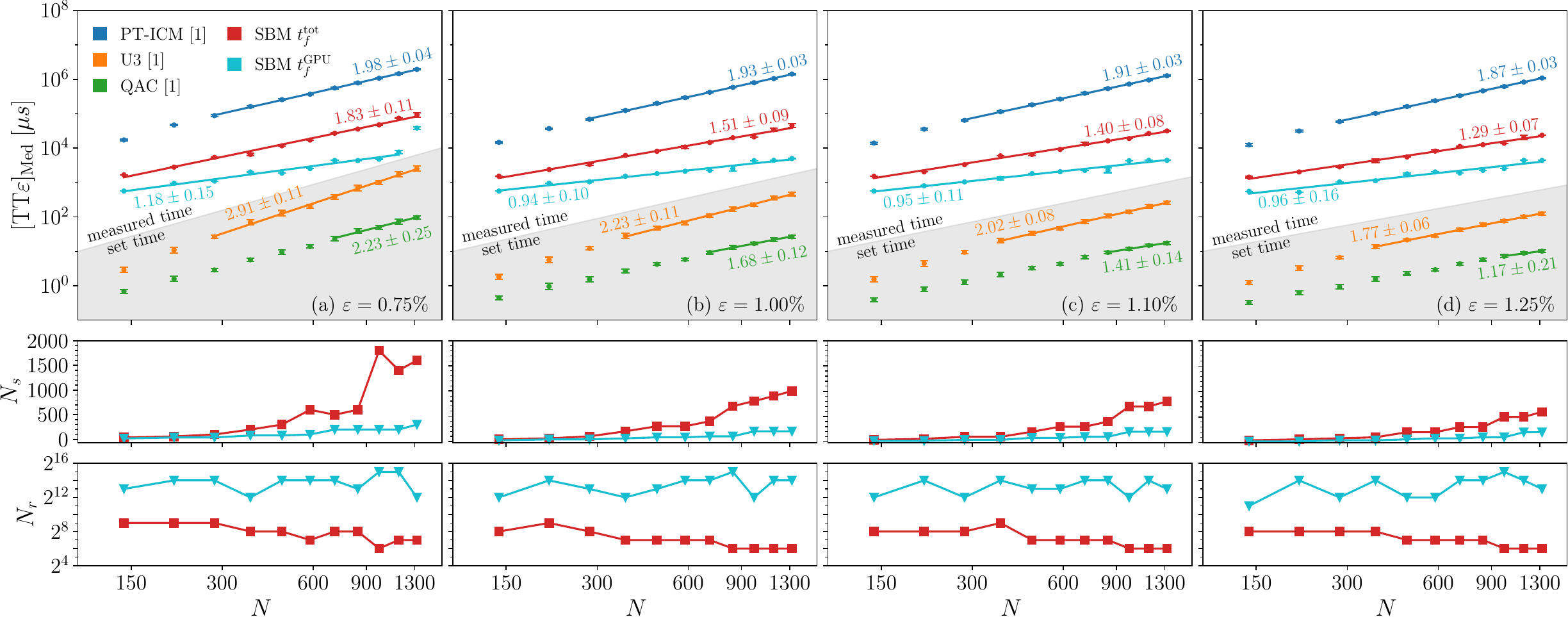}
    \caption{Time-to-$\varepsilon$ $\medTTe$ scaling with instance size $N$ for Sidon-28 problems. 
    Panels (a)--(d) correspond to target optimality gaps of
    $\varepsilon \in \{0.75, 1.00, 1.10, 1.25\}\%$. 
Data for PT-ICM (blue), U3 (orange), and QAC (green) are reproduced from Ref.~\cite{Lidar2025}. 
Remaining results concern the simulated bifurcation machine (SBM) using one GPU: 
total run-time $t_f^{\mathrm{tot}}$ (red) and pure GPU run-time $t_f^{\mathrm{GPU}}$ (cyan). 
Solid lines show power-law fits $\medTTe \propto N^{\alpha}$, with exponents $\alpha$ indicated. 
Shading differentiates set-time (shaded) from measured-time (unshaded) data. 
Even including overheads, GPU-based SBM matches or surpasses PT-ICM, U3, and QAC across all $\varepsilon$, 
challenging the notion of a quantum scaling advantage. 
Lower panels display optimal SBM parameters, number of steps $N_s$ and replicas $N_r$, minimizing $\medTTe$ 
for each $N$ and $\varepsilon$.}
    \label{fig:scaling_comp}
\end{figure*}
{\bf Thermal vs Chaotic vs Quantum Annealing --}
The scaling advantage claimed by Ref.~\cite{Lidar2025} is facilitated by the use of quantum annealing correction (QAC)~\cite{Pudenz2014}, which considerably
improves the quality of the low-energy sampling, at the cost of restricting the number and connectivity of \textit{logical} qubits.
To ensure a fair comparison, we first use exactly the same instances as in Ref.~\cite{Lidar2025}, available via Harvard Dataverse~\cite{LidarData}, and later 
extend the study to additional classes of instances.
More precisely, we consider random problems \emph{native to the topology} of the logical QAC graph
created from the D-Wave Advantage 4.1 QPU~\cite{adv4.1-properties}. The couplings are drawn from the Sidon-28 set \(\pm\{8/28,13/28,19/28,1\}\), and for each
size of the logical graph \(L \in \{5, 6, \ldots, 15\}\), we use \(125\) instances. The number of logical qubits varies
from $N=142$ ($L=5$) to $N=1322$ ($L=15$).

Following Ref.~\cite{Lidar2025}, we define the time-to-$\varepsilon$ metric as
\begin{equation}
    \label{eq:TTe}
    \TTe \doteq t_f \cdot \frac{\log(1-0.99)}{\log(1-p_{E \leq E_0 + \varepsilon |E_0|})},
\end{equation}
where \(t_f\) is the time taken to generate the sample and \(p_{E \leq E_0 + \varepsilon |E_0|}\) is the probability of finding
a solution with energy \(E\), within the \(\varepsilon\) optimality gap of the true ground state energy \(E_0\), certified by the Gurobi solver~\cite{gurobi}.
For each instance, we compute the average run-time $t_f$ and success probability 
$p_{E \leq E_0 + \varepsilon |E_0|}$ over $100$ independent solver runs. 
The median of $\TTe$ across all instances of size $N$ gives $\medTTe$, 
and its standard deviation is estimated via bootstrap resampling. 
For SBM, we distinguish two run times: the total run-time $t_f^{\mathrm{tot}}$, 
including all overhead from CPU-GPU communication, memory handling, and result aggregation, 
and the pure GPU run-time $t_f^{\mathrm{GPU}}$, representing the actual compute time 
(averaged across GPUs when applicable). 
The latter most closely corresponds to the D-Wave annealing time $\tau$, 
which omits programming and readout overheads. 
However, unlike $t_f^{\mathrm{GPU}}$, measured during execution, $\tau$ is \emph{preset} 
and cannot be independently verified afterward -- a significant distinction discussed in Sec. S5 of the Supplemental Material~\cite{supmat}.

As in Ref.~\cite{Lidar2025}, we optimize $\medTTe$ over the solver parameters, that is, number of steps $N_s$ and replicas $N_r$ (cf. Methods). 
The optimal $N_s$ and $N_r$ values are shown in the bottom rows of Fig.~\ref{fig:scaling_comp} for 
$\varepsilon \in \{0.75, 1.00, 1.10, 1.25\}\%$. 
For small instances, the larger energy spacing allows smaller $N_s$ and larger $N_r$ to enhance success probability within the optimality gap. 
As instance size grows, the Hamiltonian must vary more slowly, so typically the optimal $N_s$ increases while $N_r$ decreases, 
balancing run-time and solution quality.

\begin{figure}[htbp]
    \centering
    \includegraphics[width=0.45\textwidth]{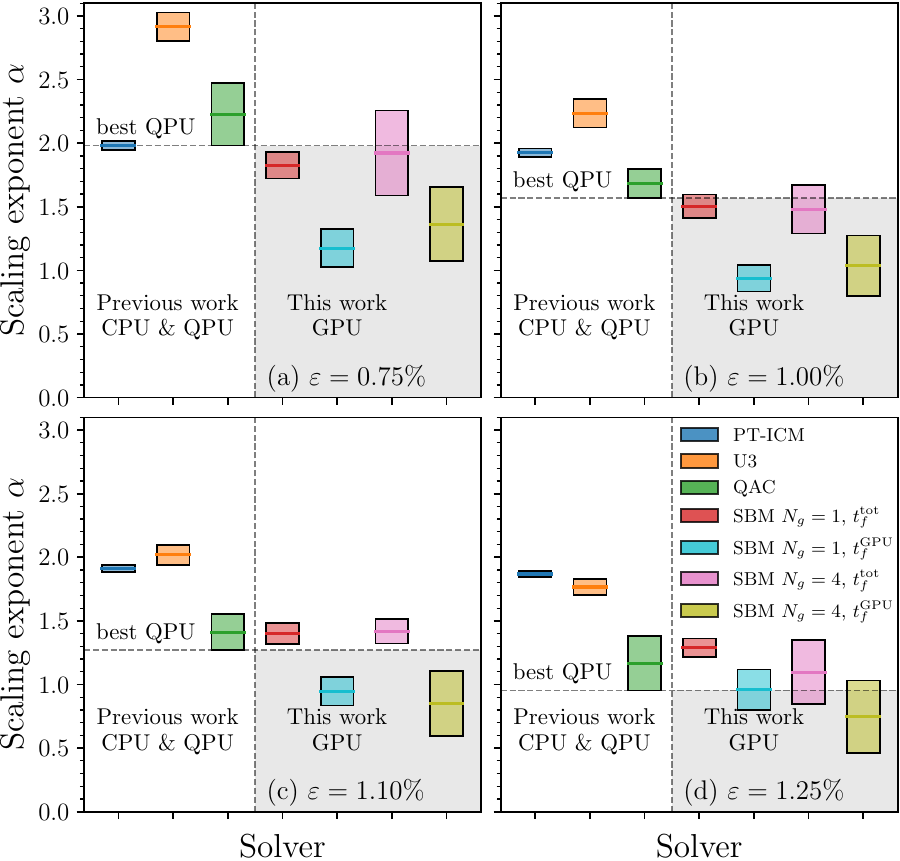}
    \caption{Time-to-$\varepsilon$ scaling exponents $\alpha$ from $\medTTe \propto N^{\alpha}$ for various solvers and 
$\varepsilon \in \{0.75, 1.00, 1.10, 1.25\}\%$ in panels (a)--(d). 
Vertical solid lines show $\alpha$ values from fits in Fig.~\ref{fig:scaling_comp}, with colored bands indicating $\pm2$ standard deviations. 
A vertical dashed line separates CPU/QPU results from Ref.~\cite{Lidar2025} (left) and our GPU-based results (right). 
The horizontal dashed line marks the best QAC fit, defining the approximate quantum-advantage boundary. 
Lightly shaded regions highlight cases closing the quantum–classical gap -- SBM $(t_f^{\mathrm{tot}}, N_g=1)$ (red, up to $\varepsilon \approx 1\%$), 
as well as $(t_f^{\mathrm{GPU}}, N_g=1)$ (cyan) and both $N_g=4$ cases (pink and light green) for all $\varepsilon$.}
    \label{fig:alpha}
\end{figure}

In the top row of Fig.~\ref{fig:scaling_comp}, we present $\medTTe$ scaling for the GPU-based SBM alongside 
the original PT-ICM, U3, and QAC results from Ref.~\cite{Lidar2025}. 
Data are shown on a log-log scale with solid lines representing power-law fits $\medTTe \propto N^{\alpha}$. 
As expected, polynomial scaling fails at $\varepsilon = 0$, since exact discrete optimization is NP-hard~\cite{Lucas2014}. 
The resulting exponents $\alpha$ are summarized in Fig.~\ref{fig:alpha}, showing that SBM with 
$N_g = 1$ and $t_f^{\mathrm{tot}}$ yields scaling comparable to, or smaller than, the best QAC results for all $\varepsilon$. 
These findings already challenge the claim of quantum scaling advantage. When considering pure GPU run-time $t_f^{\mathrm{GPU}}$ 
(analogous to the annealing time $\tau$), the gap closes further, underscoring how strongly the fitted exponents depend on overhead
in the small-instance regime. Exploiting SBM's parallelism, we also ran $N_g = 4$ GPUs, enabling larger $N_r$ without 
increasing $t_f^{\mathrm{GPU}}$; the corresponding exponents are shown in Fig.~\ref{fig:alpha} (see also Sec. S1 of the Supplemental Material~\cite{supmat}).

\begin{figure}[htbp]
    \centering
    \includegraphics[width=0.5\textwidth]{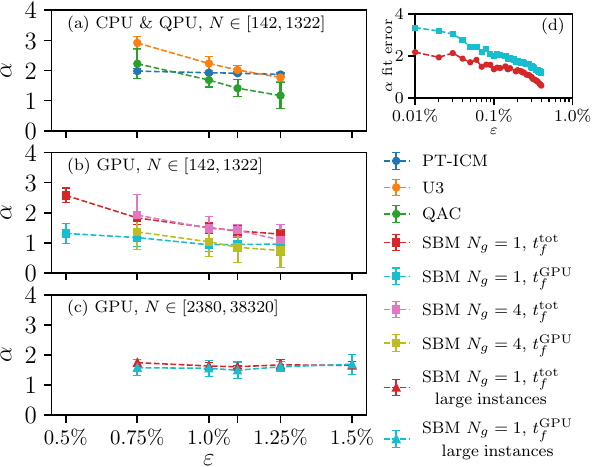}
    \caption{Dependence of the scaling exponent $\alpha$ on the optimality gap $\varepsilon$ for small (a,b) and large instances (c). 
For larger systems, scaling is more stable and overhead effects nearly vanish. 
However, $\alpha$ is expected to diverge as $\varepsilon \to 0$, marking the transition from power-law to exponential $\medTTe$ scaling. 
Panel (d) confirms this, showing the power-law fit error rising rapidly as $\varepsilon \to 0$.}
    \label{fig:alpha_vs_eps}
\end{figure}

Finally, to probe the asymptotic scaling of \(\medTTe\), we extend our analysis to systems far beyond the current capabilities of quantum
annealers, with sizes ranging from \(L=20\;(N = 2380)\) to \(L=80\;(N = 38320)\). The dependence of the scaling exponent
\(\alpha\) on the optimality gap \(\varepsilon\), extracted from these larger systems, is shown in Fig.~\ref{fig:alpha_vs_eps}(c). The scaling
becomes noticeably more robust, with overhead effects essentially vanishing. See Sec.~S2 of the Supplemental Material~\cite{supmat} for details. 

\begin{figure}[htbp]
    \centering
    \includegraphics[width=\linewidth]{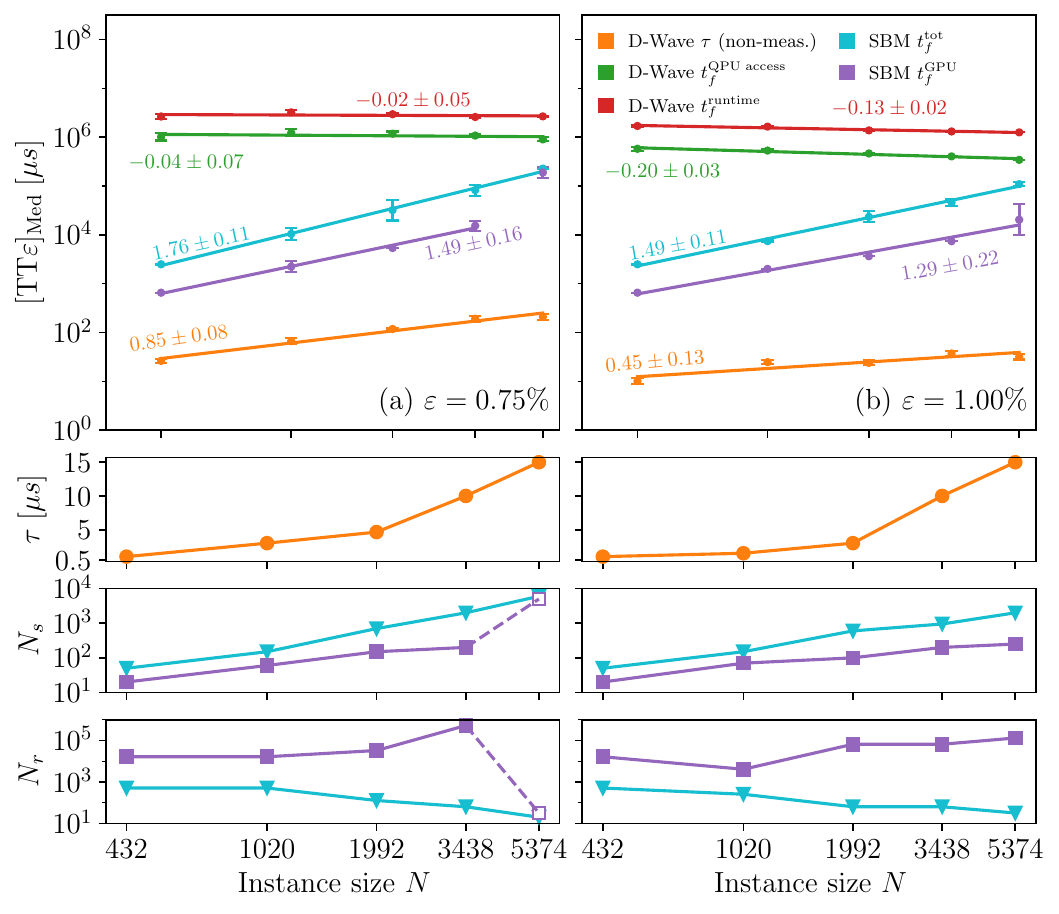}
    \caption{Time-to-$\varepsilon$ ($\medTTe$) scaling for 3D spin glasses~\cite{King2023,Chowdhury2025}, executed on the D-Wave quantum annealer 
and the SBM on a single GPU under different time-accounting schemes. Due to D-Wave's topology, embedded 3D instances 
(chains of length $2$, $\times2$ qubit overhead) were used, while SBM runs directly on the cubic lattice. 
Results for $\varepsilon \in \{0.75, 1.0\}\%$ are shown in panels (a) and (b), corresponding to the regime of nontrivial 
power-law scaling. Lower panels show optimal parameters minimizing $\medTTe$; open symbols indicate that for
given $N$, optimal value most likely lies outside the search 
range. Such points are excluded from fits. SBM results are shown for total run-time $t_f^{\mathrm{runtime}}$ (cyan) and GPU run-time 
$t_f^{\mathrm{GPU}}$ (purple). Orange, green, and red points denote annealing time~\cite{Lidar2025}, QPU access, 
and total run-time, respectively. The nonmeasurable annealing time yields smaller exponents $\alpha$, but proper time 
accounting removes this apparent advantage: for $t_f^{\mathrm{runtime}}$ and $t_f^{\mathrm{QPU\;access}}$, scaling 
exponents are near zero, indicating that asymptotic scaling is inaccessible in this size regime. Nevertheless, future devices reducing 
overhead could still realize a genuine quantum advantage.}
    \label{fig:3D_scaling}
\end{figure}

{\bf Toward quantum scaling advantage --}
A more plausible route to quantum scaling advantage is to focus on instance classes for which information can
propagate more efficiently on the annealer than in SBM-like classical dynamics. We therefore study 3D 
spin-glass instances for which the annealer attains high-quality solutions 
within the nanosecond-scale fast annealing regime~\cite{King2023,Chowdhury2025}. As shown in Fig.~\ref{fig:3D_scaling}, 
when run-time is proxied by the annealing time, $\tau$, the annealer exhibits superior scaling in the 
\(\varepsilon \leq 1\%\) regime (see also Sec. S6 of the Supplemental Material~\cite{supmat} for details). Current devices cannot yet demonstrate 
a true quantum scaling advantage, as realistic overheads (programming, readout, etc.) erase the favorable scaling. 
Nevertheless, the SBM's inability to reproduce the annealer's scaling exponent using $\tau$ as a run-time proxy 
suggests a promising path toward achieving such an advantage once hardware limitations are mitigated (cf. Sec. S5 in the Supplemental Material~\cite{supmat}).

{\bf Methods: Chaotic Dynamics --}
We consider the SBM with discretized Ising interactions governed by~\cite{Goto2021}:
\begin{equation}
    \dot{x}_i = \Delta y_i, \quad
    \dot{y}_i = -\left[\Delta - p(t)\right] x_i + \xi_0 G(x),
    \label{eq:sbm}
\end{equation}
with $G(x) = \sum_{j=1}^{N} J_{ij} f(x_j) + h_i$, where \(f(x) = \mathrm{sign}(x)\) is the sign function.
We augment the original formulation of dSB with a ternary discretization scheme~\cite{Han2023}.
Specifically, we replace $\mathrm{sign}$ function by 
$\Theta(|x| - g(t)) \cdot \mathrm{sign}(x)$, where \(g(t) = 0.7\,\frac{t}{T}\), with \(T\) the total time, and
$\Theta(x)$ the step function.

The system's nonlinearity arises from a boundary at $|x_i| = \sqrt{2}$. Exceeding this bound 
forces $x_i \rightarrow \sqrt{2}\,\mathrm{sign}(x_i)$ and sets $y_i = 0$. 
We use $\Delta = 1$ and $\xi_0 = \frac{0.7 \Delta}{\sigma \sqrt{N}}$, with $\sigma$ the standard deviation 
of the off-diagonal elements of $J$. The linear schedule $p(t) = t/T$ drives the system through a sequence of bifurcations. 
Beyond the $p(t) = p_0 = \Delta$ point, the energy landscape approximates that of the Ising Hamiltonian,
guiding convergence toward low-energy solutions, which are read out as $s_i = \mathrm{sign}(x_i)$.
The algorithm has three hyperparameters: the time step \(\Delta t\), the number of steps \(N_s\) (\(T = N_s \Delta t\)),
and the number of replicas \(N_r\). The schedule \(p(t)\) varies slowly from \(0\) to \(p_0\),
with \(N_s\) determining the slope of the ramp.
See Sec. S4 of the Supplemental Material~\cite{supmat} for additional details on the algorithm and its implementation.

{\bf Discussion $\&$ conclusions --}
\begin{figure}[htbp]
    \centering
    \includegraphics[width=0.5\textwidth]{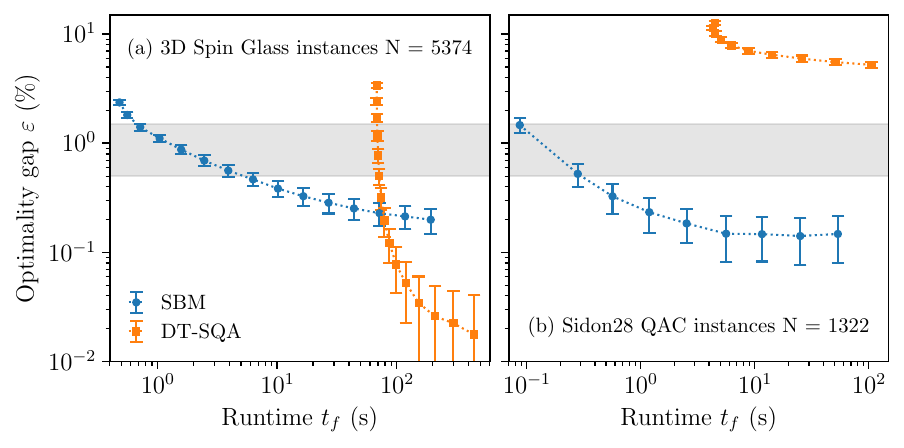}
    \caption{Average optimality gap \(\varepsilon\) achieved by SBM and DT-SQA versus measured run-time for 
(a) 3D spin-glass instances (\(N=5374\)) and (b) Sidon-28 QAC instances (\(N=1322\)). 
Averages were taken over 20 instances $\times$ 20 runs for (a) and 25 instances $\times$ 5 runs for (b). 
For Sidon-28, DT-SQA fails to reach the target range \(\varepsilon \in [0.5,1.5]\%\) (gray area), 
and thus cannot show an advantage in time-to-\(\varepsilon\) scaling. 
For 3D spin glasses, DT-SQA attains high-quality solutions but with much longer run times than SBM, 
owing to limited parallelizability~\cite{Chowdhury2025}. 
Both methods use embedded 3D instances, where the cubic lattice is minor-embedded into the Pegasus topology 
with chain length $2$~\cite{Chowdhury2025,King2023}.}
    \label{fig:gap_vs_runtime}
\end{figure}

We show that chaotic dynamics in nonlinear Hamiltonian systems, as realized in SBM,
rather than thermal fluctuations as in classical annealing, close the
quantum–classical scaling gap in approximate QUBO optimization reported in
Ref.~\cite{Lidar2025}. We further demonstrate that such quantum advantage claims
are highly sensitive to run-time definitions, rendering scaling analyses at small
system sizes ($L<20$) unreliable.

The available evidence does not permit a definitive attribution of the observed scaling to purely quantum effects. 
Our results, together with those of Ref.~\cite{Lidar2025}, support several interpretations, including that the 
D-Wave device operates in a mixed regime where classical and quantum processes coexist. In this regime, efficient 
exploration of low-energy configurations persists for $\varepsilon \leq 1.25\%$, producing scaling trends comparable, 
though not identical, to SBM-like dynamics. This is not necessarily the case for other quantum-inspired, yet classical
methods, e.g. Discrete-Time Simulated Quantum Annealing (DT-SQA)~\cite{Chowdhury2025}, cf. Fig.~\ref{fig:gap_vs_runtime}. 

In QAC, execution time is fixed rather than measured, limiting the operational meaning of $\TTe$ in Eq.~(\ref{eq:TTe}). 
Nevertheless, chaotic-like dynamics performs comparably or better even under such unfavorable conditions (cf.~Fig.~\ref{fig:alpha}). 
For larger systems, different timing definitions yield consistent results (cf.~Fig.~\ref{fig:large_instances}), confirming the 
robustness of the asymptotic behavior and providing a benchmark for future QPU studies -- though realizing this may require 
hybrid quantum-classical approaches~\cite{dwavehybrid}.

Furthermore, meaningful quantum advantage in approximate discrete optimization should be evaluated across diverse
problem instances (varying in density, size, and structure), and more importantly, as close to $\varepsilon \approx 0$ as possible. 
Relaxing the advantage criterion to sparse instances and large $\varepsilon$ values is designed to favor QPUs. However, 
SBM-like methods are naturally suited to such simulations, without requiring further modification, whereas current QPU 
architectures appear fundamentally constrained in this regime. \emph{On the problem classes and under the run-time definitions
considered here, current-generation QPUs are therefore unlikely to exhibit a clear scaling advantage over classical solvers.}

Finally, for tailored problems such as 3D spin glasses, the annealer achieves high-quality solutions within the fast annealing regime, 
exhibiting better apparent \emph{scaling} than classical methods when annealing time is used as a proxy for run-time (cf. Fig.~\ref{fig:3D_scaling}). 
These results delineate the conditions under which next-generation quantum annealers could achieve a genuine, 
though limited, scaling advantage. 

\label{sec:summary}
{\bf Acknowledgments --} This work was supported by the National Science Center (NCN), Poland, via project Sonata Bis 10, No. 2020/38/E/ST3/00269 (B.G).
Quantumz.io Sp. z o.o. acknowledges support received from The National Centre for Research and Development (NCBR), Poland, via Project
No. POIR.01.01.01-00-0061/22.

{\bf Data Availability --} The data that support the findings of this article are available from the authors upon reasonable request.
\bibliography{lit}

\newpage
\phantom{a}
\newpage
\addtocontents{toc}{\protect\clearpage}

\setcounter{figure}{0}
\setcounter{equation}{0}
\setcounter{page}{1}

\renewcommand{\thetable}{S\arabic{table}}
\renewcommand{\thefigure}{S\arabic{figure}}
\renewcommand{\thepage}{S\arabic{page}}
\renewcommand{\thesection}{S\arabic{section}}
\renewcommand{\thesubsection}{S\arabic{subsection}} 

\newcounter{scounter}
\setcounter{scounter}{0}
\renewcommand{\thescounter}{S\arabic{scounter}}

\onecolumngrid

\begin{center}
    {\large \bf Supplemental Material:\\
        Toward Quantum Scaling Advantage in Approximate Optimization \\
    }
    \vspace{0.3cm}
\end{center}
The following sections provide additional evidence supporting the robustness of SBM scaling and its connection to quantum-inspired dynamics. 
We present a detailed description of the SBM formulation and its properties, multi-GPU results, and scaling behavior in the large-instance regime. 
We also discuss runtime accounting, include additional results for the 3D spin-glass benchmarks, and provide a summary of the hardware used in this study.

\vspace{0.6cm}
\stepcounter{scounter}
\subsection{\thescounter: Multi-GPU Simulated Bifurcation Machine}
\begin{figure*}[htbp]
    \centering
    \includegraphics[width=1\textwidth]{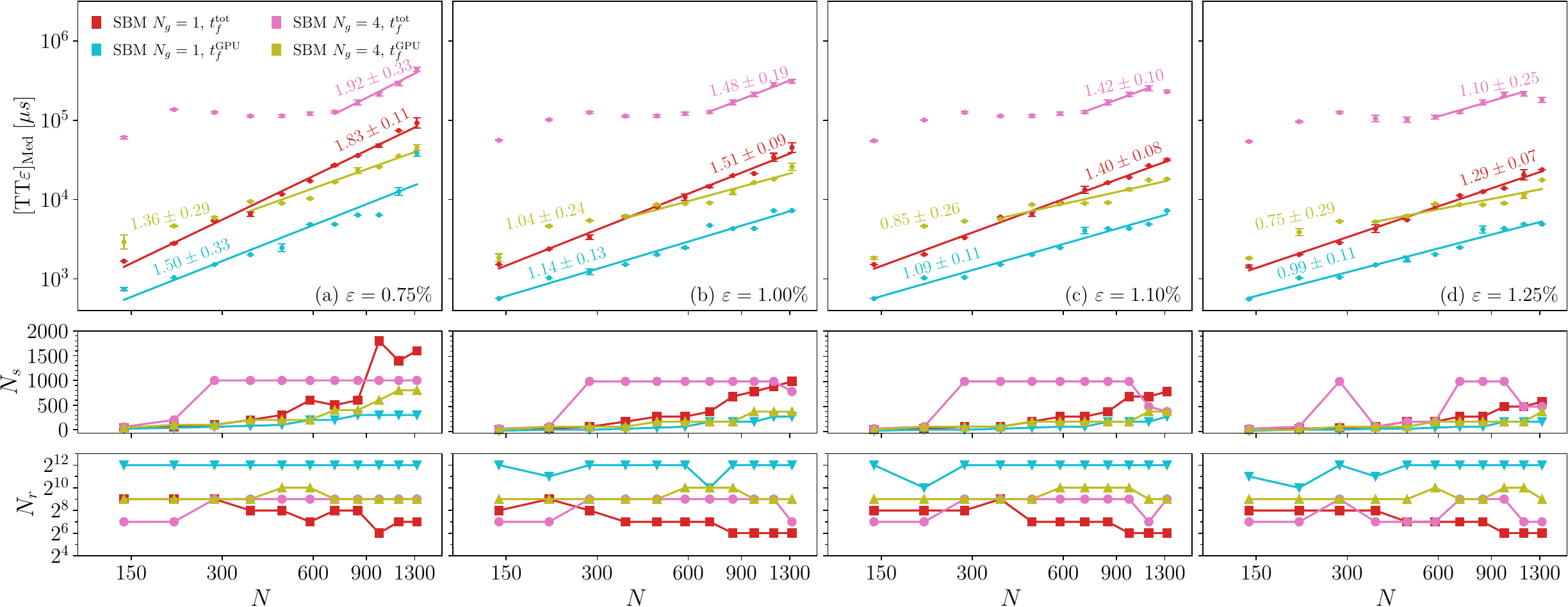}
    \caption{Time-to-$\varepsilon$ (\(\medTTe\)) scaling with the size \(N\) of the Sidon-28 instances, for values
    of \(\varepsilon \in \{0.75, 1.00, 1.10, 1.25\}\%\) in panels (a)-(d), respectively. Similar to
    Fig.~\ref{fig:scaling_comp} in the main text, but comparing the SBM performance for \(N_g=1\) (red, cyan)
    and \(N_g=4\) (pink and light green). Despite the visible impact of the increased overhead associated with
    the multi-GPU implementation, the (tentative) scaling properties are improved. This highlights one of the
    advantages of the SBM algorithm, since additional GPUs can be added to improve the performance, without
    the need for a generational hardware upgrade.}
    \label{fig:multi_gpu}
\end{figure*}
In the main text, we have discussed the nature and details of the Simulated Bifurcation Machine (SBM) algorithm,
and how it can be used to solve QUBO/Ising problems. One of the core benefits of replacing thermal fluctuations
with chaotic dynamics is the inherent parallelizability of the algorithm, which allows for efficient use of modern
hardware accelerators, such as GPUs.

Here, we present a tentative comparison of scaling properties for \(N_g=1\) and \(N_g=4\) GPUs. The results are shown
in Fig.~\ref{fig:multi_gpu} where, as in the main text, we plot the time-to-$\varepsilon$ scaling with the instance
size \(N\). The scaling properties demonstrate an improvement, with the scaling exponent \(\alpha\) being smaller
for \(N_g=4\) than for \(N_g=1\), when compared for the same type of runtime measurements.
However, an increased number of GPUs comes at the cost of additional overhead associated with dispatching and
collecting the computations from multiple GPUs, whose impact is especially pronounced for problems as small as those considered here.
This is reflected in the results for \(N_g=4\) and \(t^{\mathrm{tot}}_f\) (pink), with individual values of \(\medTTe\)
being larger by a factor of \(\sim 100\) when compared to the \(N_g=1\) case. Of course, we expect the impact of the
overhead to diminish with increasing instance size. Moreover, it is possible to disentangle the overhead from the actual
computation time, and obtain the pure GPU computation time \(t^{\mathrm{GPU}}_f\) (light green). Values of \(\medTTe\)
obtained in this way maintain the improved scaling, yet are comparable to the ones obtained for \(N_g=1\).

\begin{figure*}[htbp]
    \centering
    \includegraphics[width=1\textwidth]{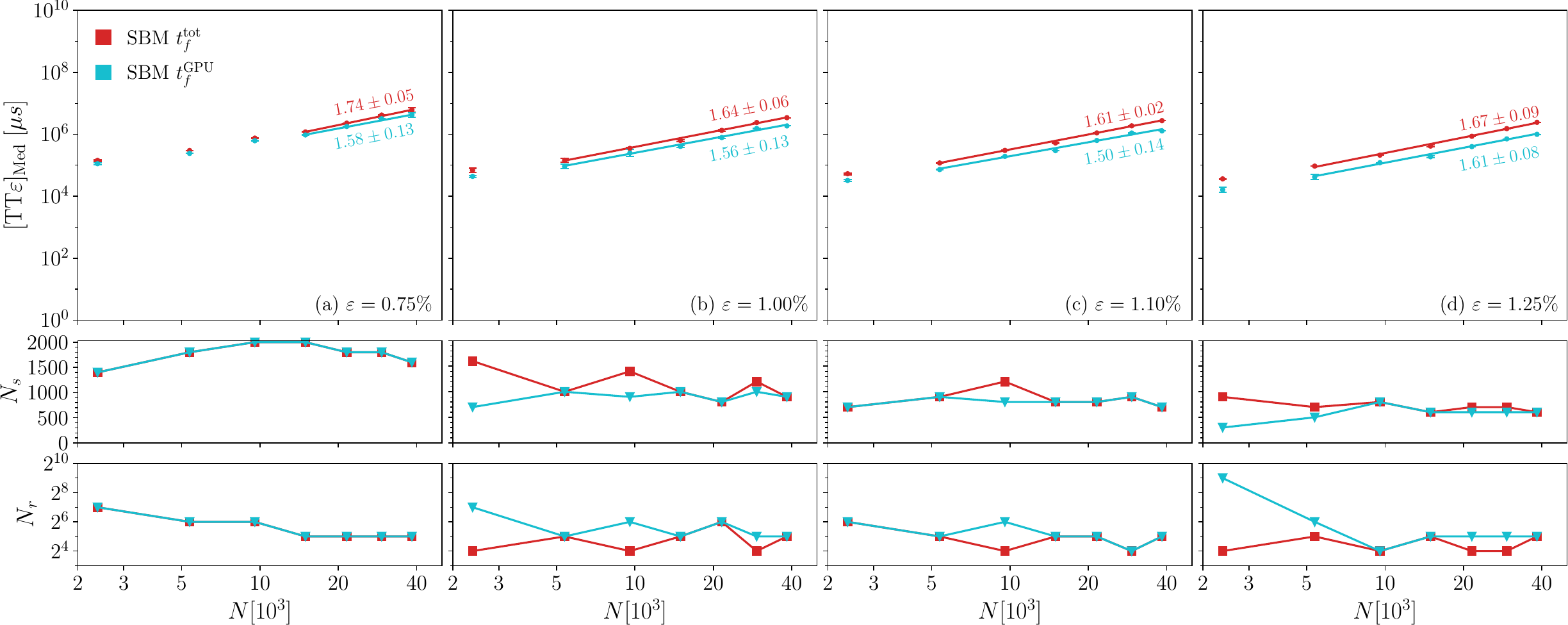}
    \caption{Time-to-$\varepsilon$ ($\medTTe$) scaling with the size \(N\) of the Sidon-28 instances, for values of
    \(\varepsilon \in \{0.75, 1.00, 1.10, 1.25\}\%\) in panels (a)-(d), respectively. Results are in the range of
    instance sizes \(N \in [10^3, 10^4]\), far beyond the current capabilities of quantum annealers. As expected,
    the impact of the overhead (\(t^{\mathrm{tot}}_f\) vs \(t^{\mathrm{GPU}}_f\)) becomes negligible in this regime,
    indicating that the asymptotic scaling is indeed dominated by the actual computation time.}
    \label{fig:large_instances}
\end{figure*}
\stepcounter{scounter}
\subsection{\thescounter: Scaling in the large instance regime}
The range of instance sizes discussed in Ref.~\cite{Lidar2025} and in the main text is limited by the current capabilities
of quantum annealers. The Advantage series QPUs have an underlying topology of a \(P_{16}\) graph, which limits the QAC
procedure to logical graphs with \(L\leq 15\), corresponding to a number of logical qubits \(N \leq 1322\). SBM, on the
other hand, is free of such limitations, and can be used to solve problems of arbitrary topology and size, limited only
by the available memory. Therefore, we can extend the scaling analysis to much larger instances and move a step closer
to the regime that allows for drawing conclusions about the asymptotic properties.

To this end, we consider the same Sidon-28 instances, but with sizes ranging from \(L=20\;(N = 2380)\) to \(L=80\;(N = 38320)\),
which is more than an order of magnitude larger than the largest instances considered in the main text.
Performing experiments with a quantum annealer in this regime would require a Pegasus topology QPU with at
least \(1.5\cdot 10^5\) physical qubits. If the current growth rate in qubit counts is maintained, such
a device could become available in the next \(5\) to \(10\) years~\cite{veloxq2025}.
For each \(L\), we consider \(10\) random instances, and estimate success probability and average runtime
from \(100\) independent shots. The results are shown in Fig.~\ref{fig:large_instances}, where we again
plot the time-to-$\varepsilon$ as a function of the instance size \(N\), for the same values of \(\varepsilon\) as before.
First and foremost, we observe that both the scaling exponents and optimal values of parameters exhibit much weaker
dependence on the strategy of time measurements, indicating a strongly diminished impact of the overhead. This is expected,
and hints that we are indeed closer to the \textit{robust} asymptotic regime, where the actual computation time dominates
the overall runtime. 

\stepcounter{scounter}
\subsection{\thescounter:  Dequantization of Quantum SBM: Detailed analyses}
We begin with a detailed derivation of the equations underlying the SBM
algorithm~\cite{Goto2019}. The starting point of this discussion is a single
parametrically driven Kerr nonlinear oscillator (KPO) with a time-dependent
Hamiltonian (units with \(\hbar = 1\) are assumed): 
\begin{equation}
  H(t) = \Delta a^\dagger a + \frac{K}{2} a^{\dagger 2}a^2 - \frac{p(t)}{2}(a^{\dagger 2} + a^2), 
\end{equation}
where \(a,a^\dagger\) are the bosonic annihilation and creation operators respectively,
satisfying canonical commutation relations \([a,a^\dagger] = 1\), and the drive
\(p(t) = p_{\text{max}} t/T\) is a linear ramp from 0 to \(p_{\text{max}}\),
with constant \(T\) controlling the slope. Other parameters are detuning
\(\Delta\) and Kerr nonlinearity strength \(K\)~\cite{Goto2019}. 

Initially, for \(p(t)=0\), the ground state is the vacuum \(\ket{\psi(0)}=\ket{0}\). 
As \(p\) increases, the system follows the ground state adiabatically. 
For \(p(t)\gg\Delta\), it approaches the ground state of \(H(t)\) with \(\Delta=0\), 
a superposition of two coherent states arising from parity symmetry:
\begin{equation}
  \ket{\psi(t)} = \frac{1}{\sqrt{2}}\left(\ket{\sqrt{p/K}} + \ket{-\sqrt{p/K}}\right),
\end{equation}
where the overlap \(\braket{\sqrt{p/K}}{-\sqrt{p/K}} = \mathrm{e}^{-p/2K}\) is
exponentially small for large \(p\) and can be ignored. This is known as a 
Schrödinger cat state. The ground state has this form because, for \(\Delta = 0\),
the Hamiltonian can be written as, up to a constant term, \(H = K/2(a^{\dag 2} -p/K)(a^2 - p/K)\).
Our aim is to use a network of KPOs to solve Ising spin problems
(for simplicity we omit local fields - linear term):
\begin{equation}
  E_{\text{Ising}} = - \sum_{i<j} J_{ij}s_i s_j \quad (s_i = \pm 1).
\end{equation}
To this end, consider an \(N\)-oscillator network that is coupled via linear
interaction
\begin{equation}
  H_N(t) = \sum_{i=1}^N H^{(i)}(t) - \frac{\xi_0}{2} \sum_{i,j} J_{ij}(a_i^\dagger a_j + a_i a_j^\dagger),
  \label{eq:network_hamiltonian}
\end{equation}
where \( H^{(i)}(t) \) is the Hamiltonian of the \(i\)-th KPO. Detunings \(\Delta_i\) are tuned to ensure the vacuum state \(\ket{0}\) is the initial 
ground state. More precisely, this requires the following matrix \(M\)
\begin{equation}
  M_{ii} = \Delta_i, \quad M_{ij} = -\xi_0 J_{ij} \quad (i \neq j),
\end{equation}
to be positive semidefinite. An appropriate choice is \(\Delta_i =  \xi_0 \sum_j
\abs{J_{ij}}\), cf. Ref.~\cite{Goto2016}. Post bifurcation, the ground state of
the system of \(N\) KPOs is \(2^N\)-fold degenerate, and spanned by the tensor
products of cat states \(\ket{\pm\sqrt{p/K}}\). The Ising interactions
introduce a perturbative correction to the energy of each tensor product
\(\ket{s_1 \sqrt{p/K}} \otimes \cdots \otimes \ket{s_N \sqrt{p/K}}\):
\begin{equation}
  E_{\text{corr}}(\{s_i\}) \propto -\xi_0 \sum_{i,j} J_{ij}s_i s_j,
\end{equation}
which lifts the exponential degeneracy in favor of the doubly degenerate state
with minimal \(E_{\text{Ising}}\). Taking into account the parity symmetry
(that is \(a_i \to -a_i\)), the final state after the adiabatic
evolution takes the form
\begin{equation}
  \ket{\psi_{\text{final}}} = \frac{\ket{s_1 \sqrt{p/K}} \otimes \cdots \otimes 
  \ket{s_N \sqrt{p/K}} + \ket{-s_1 \sqrt{p/K}} \otimes \cdots \otimes \ket{-s_N \sqrt{p/K}}}{\sqrt{2\left(1 + \mathrm{e}^{-2Np/K}\right)}}.
\end{equation}

Let us now discuss the construction of a classical system inspired by the above
quantum system. To this end consider the Heisenberg equation of motion for a
single oscillator mode
\begin{equation}
  \dot{a} = i[H, a] = i \left( p a^\dag - \Delta a - K a^\dag a^2 \right).
  \label{eq:heisenberg_EOM}
\end{equation}
During adiabatic evolution, the system is at all times in an approximately
coherent state and the moments can be approximated \(\langle a^{\dag m}
a^n\rangle \approx \alpha^{\ast m} \alpha^n\). Taking the expectation value of
Eq.~\eqref{eq:heisenberg_EOM}, using \(\alpha = x + iy\) and separating
equations into the real and imaginary parts, we obtain
\begin{align}
  \dot{x} &= \left[\Delta + p + K(x^2 + y^2)\right] x = \pdv{\mathcal{H}}{y}, \\
  \dot{y} &= \left[-\Delta + p - K(x^2 + y^2)\right] y = -\pdv{\mathcal{H}}{x},
\end{align}
where \(\mathcal{H} = \frac{-p}{2} \left(x^2-y^2\right) + \frac{\Delta}{2}
\left(x^2+y^2\right) + \frac{K}{4} \left(x^2+y^2\right)^2\). This model exhibits
a pitchfork bifurcation at \(p = \Delta\), where the stable fixed point at the
origin \((x,y)=(0,0)\) becomes unstable and two new stable fixed points appear,
located at \((\pm \sqrt{(p-\Delta)/K}, 0)\). Since the fixed points \(\dot{x} =
\dot{y} = 0\) coincide with the minima of the Hamiltonian \(\mathcal{H}\), and
for a fixed \(p\) the dynamics are energy-conserving, this bifurcation can be
understood in terms of an adiabatic invariant, $I = \oint y \dd{x}$, 
which corresponds to the area enclosed by the trajectory in the phase space.
During the adiabatic evolution, this invariant stays constant, so the
trajectories that initially closely encircle the origin, after the bifurcation
will closely encircle the new stable fixed points~\cite{Goto2019}.

We can extend this model to a network of \(N\) oscillators by considering \(N\)
modes \(a_i\) coupled via the Ising interaction term, by following the same
procedure of taking the expectation value of Heisenberg EOM in a coherent state,
and separating into real and imaginary parts. This leads to
\begin{align}
    \dot{x}_i &= y_i\left[\Delta_i + p + K(x_i^2 + y_i^2)\right] - \xi_0 \sum_j J_{ij} y_j, \\
    \dot{y}_i &= x_i\left[-\Delta_i + p - K(x_i^2 + y_i^2)\right] + \xi_0 \sum_j J_{ij} x_j.
\end{align}
Since the momenta vary around zero during the dynamics, some terms proportional
to \(y_i\) can be neglected, leading to the simplified equations
\begin{align}
    \dot{x}_i &= \Delta y_i = \pdv{\mathcal{H}_{\text{SBM}}}{y_i}, \\
    \dot{y}_i &= \left[-\Delta_i + p - K x_i^2\right] x_i + \xi_0 \sum_j J_{ij} x_j = -\pdv{\mathcal{H}_{\text{SBM}}}{x_i},
\end{align}
where we assumed \(\Delta_i = \Delta\) for all oscillators. We arrive at
Hamiltonian equations of motion, with the Hamiltonian:
\begin{equation}
    \mathcal{H}_{\text{SBM}} = \sum_i \frac{\Delta}{2} y_i^2 + 
    \sum_i \left[\frac{\Delta-p}{2} x_i^2 + \frac{K}{4} x_i^4 \right] - \frac{\xi_0}{2} \sum_{i,j} J_{ij} x_i x_j.
\end{equation}
These equations can be further simplified by placing non-elastic walls at
$\abs{x_i}=\sqrt{2}$, dropping the fourth-order terms from the potential and introducing a
ternary discretization scheme. The final equations of the SBM are
\begin{align}
    \dot{x}_i &= \Delta y_i = \pdv{\mathcal{H}_{\text{SBM}}}{y_i}, \\
    \dot{y}_i &= \left[-\Delta_i + p\right] x_i + \xi_0 \sum_j J_{ij} f(x_j) = -\pdv{\mathcal{H}_{\text{SBM}}}{x_i}, 
\end{align}
with $f(x) = \Theta(|x| - g(t)) \cdot \mathrm{sign}(x)$ and \(g(t) = 0.7\,\frac{t}{T}\).
In addition, when $\abs{x_i}>\sqrt{2}$, the update $x_i = \sqrt{2}\,\mathrm{sign}(x_i)$, $y_i = 0$ is applied.

The introduction of the walls prevents the system trajectories
from becoming unstable and divergent, which sometimes occurs during the evolution of the system with the fourth-order potential term.
See Alg.~\ref{alg:sbm} for a pseudocode description of the SBM algorithm.

\begin{figure}[htbp]
    \refstepcounter{algorithm}\label{alg:sbm}
    \textbf{ALG.~\thealgorithm.} Simulated Bifurcation Machine.\par
    \vspace{-0.7em}\rule{\linewidth}{0.6pt}\vspace{-0.3em}\par
    \begin{algorithmic}\renewcommand{\baselinestretch}{1.25}\selectfont
        \State{\textbf{Input:} \\
            \qquad $J,\, h$ -- coupling matrix and local fields vector specifying an instance of the Ising model, \\
            \qquad $\Delta t$ -- time step, \\
            \qquad $N_s$ -- number of steps, \\
            \qquad $p(t)$ -- pump function, \\
            \qquad $f(x)$ -- discretization function, \\
            \qquad $\Delta,\, \xi_0$ -- hyperparameters. }
        \State{\textbf{Output:} \\
            \qquad $E,\, q$ -- energy and its corresponding state of the found minimum.}
        \State{$n = \mathrm{length}(h)$}
        \State{$x_1 \leftarrow \{\text{rand}(-1,1)\}^n$}
        \State{$y_1 \leftarrow \{\text{rand}(-1,1)\}^n$}
        \For{$i=1,\,\ldots,\,N_s$}
        \State{$y_{i+1} \leftarrow y_i + \left\{ \left[p(j \Delta t) - \Delta\right]  x_i+ \xi_0 \left(\sum_{j=1}^{N} J_{ij} f(x_j) + h_i\right) \right\} \Delta t $}
        \State{$x_{i+1} \leftarrow x_i + \Delta \cdot y_{i+1} \cdot \Delta t$}
        \For{$j=1,\,\ldots,n$}
        \If{$\abs{(x_{i+1})_j} \geq \sqrt{2}$}
        \State{$(x_{i+1})_j \leftarrow \sqrt{2}\,\mathrm{sign}((x_{i+1})_j) $}
        \State{$(y_{i+1})_j \leftarrow 0 $}
        \EndIf
        \EndFor
        \EndFor
        \State{$x = \mathrm{sign}(x_{N_s+1})$}
        \State{$E = -\frac{1}{2} x^T J x - h^T x$}
        \State \Return $E,x$
    \end{algorithmic}
    \vspace{-0.6em}\rule{\linewidth}{0.6pt}
\end{figure}
\stepcounter{scounter}
\subsection{\thescounter: How SBM selects good quality solutions.}
\begin{figure}[htbp]
    \centering
    \includegraphics[width=0.7\textwidth]{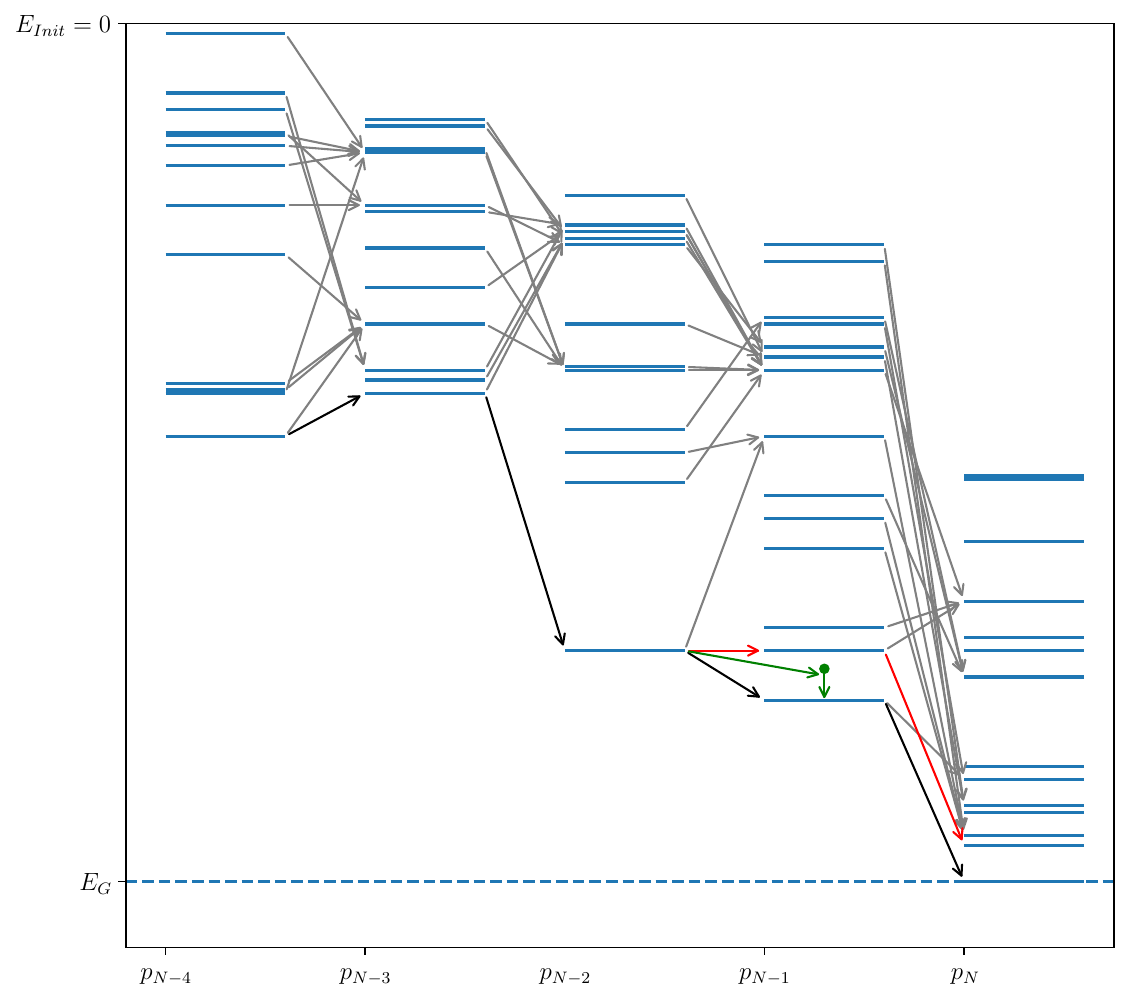}
    \caption{
    A schematic illustration of the evolving energy landscape and its exploration by the SBM algorithm.
    Black arrows denote a chain of bifurcations in which, at each bifurcation, the lowest-energy minimum
    maps to a new lowest-energy minimum. If this behavior persists throughout the evolution,
    the adiabatic dynamics leads to the ground state. Red arrows illustrate a scenario in
    which the minimum at bifurcation $N-1$ evolves into a higher-energy local minimum, preventing
    the system from reaching the ground state. Green arrows indicate a saddle-node bifurcation at $N-1$,
    where the minimum disappears; the intrinsic dynamics may then relax the system into a new lowest-energy minimum,
    ultimately reaching the ground state.
    }
\label{fig:bifurcations}
\end{figure}

Let us explain why the SBM Hamiltonian encodes the ground state of an Ising model instance, specified by $J_{ij}$. 
Our focus is on the potential energy, defined as 
\begin{equation}
  U(x) = \sum_i \left[\frac{\Delta-p}{2} x_i^2 + \frac{K}{4} x_i^4 \right] - \frac{\xi_0}{2} \sum_{i,j} J_{ij} x_i x_j.
\end{equation} 
We also denote the Hessian matrix as
\begin{equation}
  H(x)_{ij} = \frac{\partial^2 U(x)}{\partial x_i \partial x_j} = \left[3 K x_i^2 +\Delta - p  \right] \delta_{ij} - \xi_{0} J_{ij}. 
\end{equation}
For a small value of $p$, such that  $\Delta-p>0$, the energy exhibits one stable minimum at $x_i = 0$ for all $i$. This minimum loses its stability when
the value of $p$ is increased, and crosses $\Delta + \lambda_{\text{Max}}$, where $\lambda_{\text{Max}}$ is the maximal eigenvalue corresponding to the 
maximal eigenvector of $J$~\cite{Wang2023,CIMStatPhys1,CIMStatPhys2}. This is the first bifurcation that leads to two new minima, whose location is 
related to the $J$ maximal eigenvector. As the value of $p$ increases further, a subsequent sequence of bifurcations occurs. The bifurcations are associated
with the creation of new critical points (at which energy gradient $\grad_x U(x)$ vanishes). For $p \gg 0$ the Ising energy term can be treated as 
a small perturbation, and the perturbation theory can be used to show that~\cite{CIMStatPhys1,Goto2016}
\begin{equation}
  U(x) = -\frac{N}{4} (1-r) \left( \frac{p-\Delta}{K} \right)^2 + \frac{p-\Delta}{2K} \sum_{ij}J_{ij} \sigma_i \sigma_j + O(1), \quad \sigma_i = \text{sgn}(x_i),
\end{equation} 
where $N$ is the total number of spins, and $r$ is the number of spins for which $x_i = 0$ (note that such spins also do not contribute to the Ising energy term). 
Thus, for large $p$, the sign configuration of the ground state of SBM corresponds to the ground state of the Ising energy instance.
However, there is no guarantee that adiabatic evolution, in which the system stays in the minimum energy state after each bifurcation, would lead to the final
minimum energy state. This is because at each bifurcation there are three possibilities (cf. Fig.~\ref{fig:bifurcations})~\cite{CIMStatPhys1,CIMStatPhys2}:
\begin{enumerate}
  \item The pre-bifurcation minimum is mapped onto the post-bifurcation minimum. If this correspondence holds for each bifurcation in the sequence, the annealing process would 
  lead to the global minimum, and a theorem analogous to that used in adiabatic quantum computing could be formulated~\cite{AdiabaticQCReview}. 
  The energy gap would be defined as the minimum gap between the lowest-energy state and the first excited state at each bifurcation.  
  \item The pre-bifurcation minimum is mapped onto a higher-energy post-bifurcation minimum. In this case, the global minimum might not be reached,
  depending on the subsequent evolution of this minimum. This situation is somewhat
  analogous to quantum annealing failing to reach the ground state due to Landau–Zener–like transitions occurring at avoided level crossings~\cite{PhasesQA}.
  \item A saddle-node bifurcation occurs, causing the pre-bifurcation minimum to disappear; the dynamics may then settle into another local minimum. The final ground state can still
  be reached, depending on how the system evolves toward the minimum determined by relaxation. Note that such a mechanism is absent in quantum annealing, where the 
  evolution is strictly unitary. Only the inclusion of additional, uncontrollable degrees of freedom -- usually referred to as the environment -- can induce relaxation 
  from a higher to a lower energy level. 
\end{enumerate}
Therefore, to reach the ground state with certainty, one would need to analyze the entire evolution of the energy landscape during the drive and identify a continuous 
path connecting the initial local minimum to the final ground state. However, because the bifurcation dynamics are instance-dependent, such an approach is not 
feasible as a general algorithmic principle. 
Similarly, in quantum annealing, the minimum gap is not known \emph{a priori}, and the optimal annealing time necessary to reach 
the ground state with high probability cannot be determined in advance. Instead, one typically samples low-energy states for various annealing times and selects 
the best solution~\cite{AdiabaticQCReview}. Likewise, in the SBM, taking advantage of its intrinsic parallelism, one can explore the energy landscape by initializing multiple trajectories 
from different initial conditions and observing how they evolve through successive bifurcations.

In the numerical implementation, the fourth-order potential term is replaced by inelastic walls at 
$x_i = \pm \sqrt{2}$, and a discretization scheme is applied. Although analytical treatment is not possible 
due to the non-analytic nature of the walls, the system's behavior in the large-pump regime remains 
qualitatively similar. The walls constrain the coordinates to two values, effectively rendering the system 
binary. In this regime, the discretization does not alter the Ising energy term, which remains a small 
perturbation depending on the system state. 
\stepcounter{scounter}
\subsection{\thescounter: Runtime accounting}
\begin{figure}[htbp]
    \centering
    \includegraphics[width=1\textwidth]{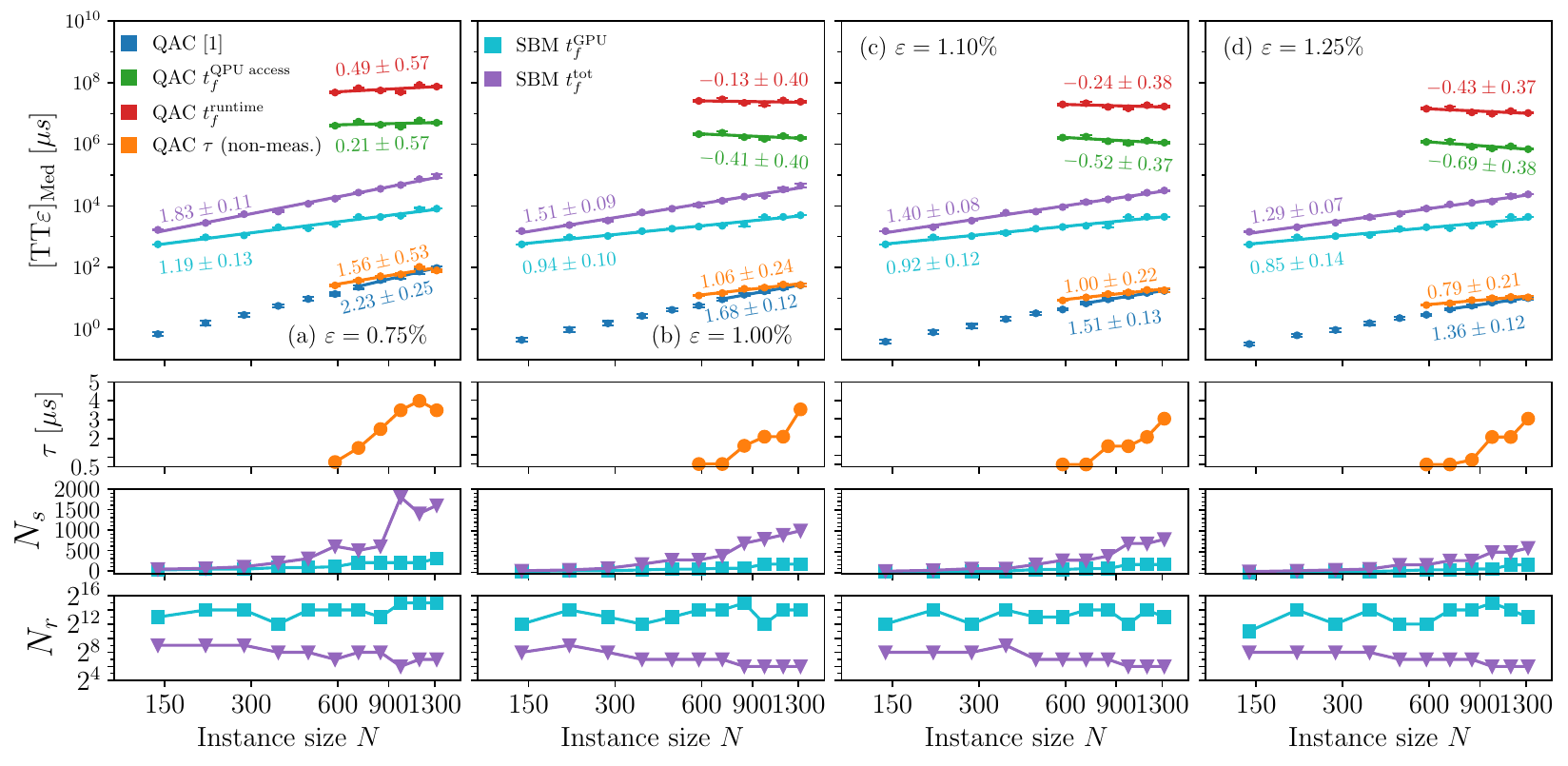} 
    \caption{
        Time-to-$\varepsilon$ ($\medTTe$) scaling for the S28 QAC instances from the main text, executed on both the D-Wave 
quantum annealer and the SBM running on a single GPU under different time-accounting schemes. Proper time accounting, including 
overheads, significantly affects the scaling analysis. Orange data points (and blue, reproduced from Ref.~\cite{Lidar2025}) 
use annealing time per sample as a runtime proxy. This time is set by the end user before execution and does not necessarily 
reflect the actual computation time on hardware. Since the D-Wave device does not return effective annealing duration, we refer 
to this quantity as ``non-measurable.'' Bottom rows show the optimal values of parameters
over which the time-to-$\varepsilon$ is minimized.}
    \label{fig:scaling_s28}
\end{figure}
We begin with two important definitions. 
The \emph{runtime}, $t$, is a measurable quantity representing the \textbf{actual duration} of computation 
(as one would measure on hardware). In contrast, the \emph{annealing time}, $\tau$, is a \textbf{user-defined control parameter} 
specifying how long the annealing process is \emph{intended} to run, set prior to execution. Ideally, $t(N) \approx \tau(N) + C$, 
where $C$ is an overhead term that may weakly depend on the problem size $N$. Crucially, the condition 
$C \ll t(N_{\rm final}) - t(N_{\rm initial})$ should hold, where $N_{\rm initial}$ and $N_{\rm final}$ 
are the smallest and largest problem sizes in the scaling analysis. When satisfied, a nonzero $C$ only 
shifts scaling curves slightly without affecting the fitted exponents, which is why $C$ is often omitted in scaling analyses.

With these definitions in place, we can examine how deviations from this assumption affect scaling behavior.
To this end, we performed additional simulations, whose results are shown in Fig.~\ref{fig:scaling_s28}. 
The most general timing metric is the external runtime, $t_f^{\mathrm{runtime}}$, which includes all operational 
times of the QPU, queuing delays, and communication overhead between the user's system and the QPU. 
This is the only measurement that can be unambiguously and independently verified without direct hardware access. 
Next, we define the QPU access time, $t_f^{\mathrm{QPU\;access}}$, which excludes communication overhead but includes 
all operational components of a single QPU job~\cite{dwave_operation_timing} -- programming, postprocessing, annealing, 
readout, and thermalization per sample. Finally, we consider the annealing time per sample, $\tau$, as used in 
Ref.~\cite{Lidar2025}. Our analysis shows that, for timing measurements of practical relevance, 
$C \gg t(N_{\mathrm{final}}) - t(N_{\mathrm{initial}})$ for current annealers, rendering the fitted scaling 
exponents statistically unreliable.

For the SBM simulations, the dominant overhead arises from CPU–GPU data transfer 
before integrating the equations of motion.
We denote by $t_f^{\mathrm{tot}}$ the total time including this overhead,
(analogous to $t_f^{\mathrm{QPU\; access}}$), while the time spent in the
main GPU compute loop is denoted as $t_f^{\mathrm{GPU}}$ (analogous to $\tau$).
We note that \(\tau\) 
on quantum annealers refers to expected annealing time \emph{per sample}, and multiple samples are 
processed sequentially. In contrast, SBM integrates the equations of motion for all samples in 
parallel, so \(t_f^{\mathrm{GPU}}\) corresponds to the total compute time for all samples.
Other initialization overheads, common to both classical and quantum platforms, are neglected. 
Since these sources are well understood, robust scaling for large $N$ is observed, 
independent of runtime definition, as shown in Fig.~\ref{fig:large_instances}\dots
For small $N$, scaling depends more sensitively on the runtime definition, yet overall conclusions remain unchanged.
This indicates that for SBM we have $C \ll t(N_{\rm final}) - t(N_{\rm initial})$, as desired. 

In contrast, for quantum annealers, an apparent scaling trend emerges only when 
annealing time is used as a runtime proxy. Even under the favorable assumption $\tau \approx t$, 
the quantum hardware does not exhibit any form of “advantage” relative to a well-optimized 
quantum-inspired algorithm such as SBM running on a single GPU.

\begin{figure}[htbp]
    \centering
    \includegraphics[width=1\textwidth]{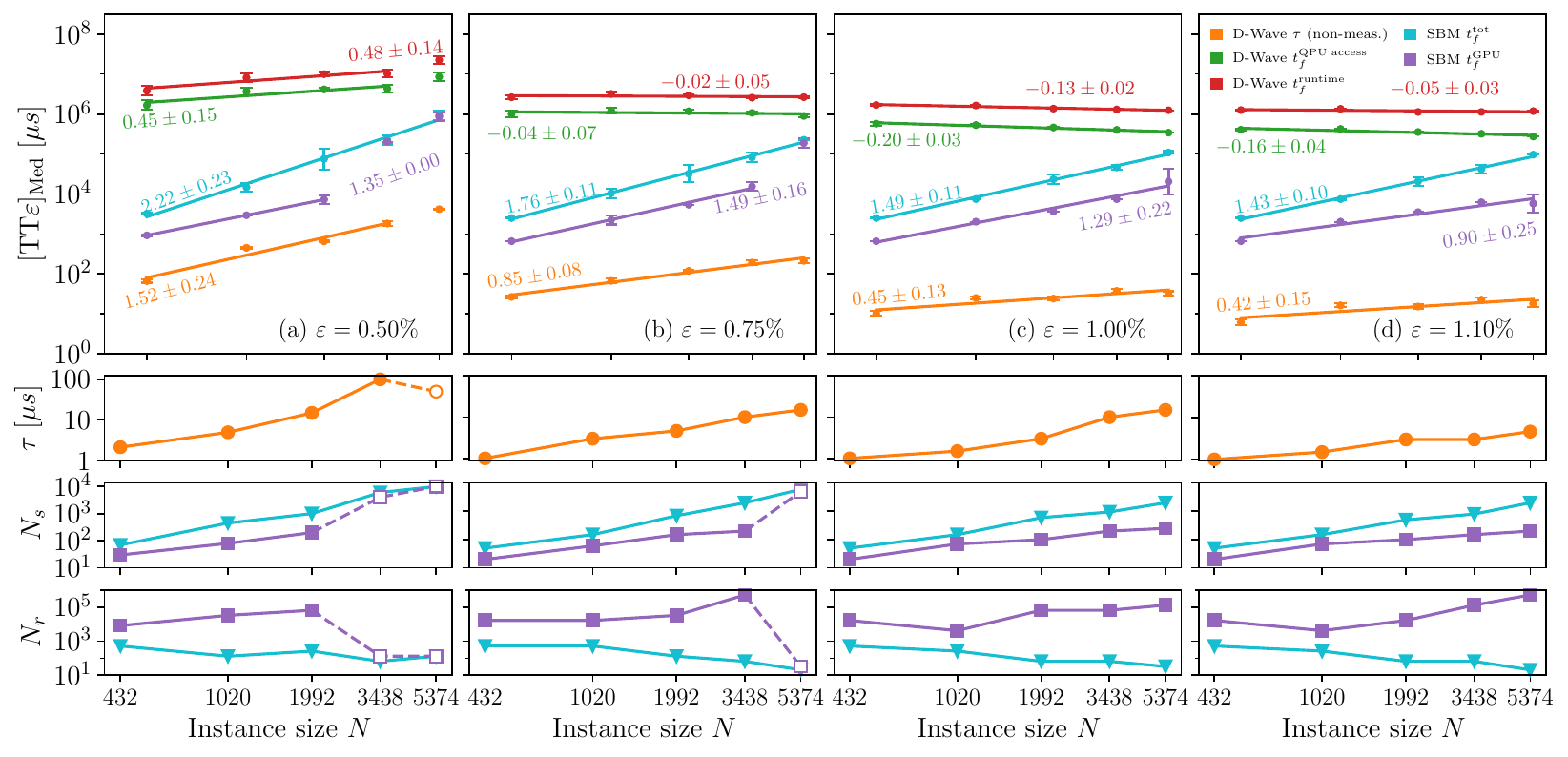} 
    \caption{Time-to-$\varepsilon$ ($\medTTe$) scaling for 3D spin glasses~\cite{King2023,Chowdhury2025}, obtained on the D-Wave quantum annealer 
and the SBM on a single GPU under different timing schemes. Panels (b,c) reproduce Fig.~4 from the main text, while (a,d) extend 
the results to $\varepsilon = 0.50\%$ and $1.10\%$. Lower panels show optimal parameters minimizing $\medTTe$; open symbols mark 
values outside the explored range. For $\varepsilon = 0.50\%$, $\medTTe$ for the largest instances cannot be reliably minimized 
for either solver, leading to a rapid increase that signals the onset of exponential rather than power-law scaling. 
For $\varepsilon = 1.10\%$, both SBM and D-Wave maintain robust power-law behavior, with the quantum annealer's exponent 
remaining close to that at $\varepsilon = 1.00\%$, while its optimal annealing time decreases notably. 
This indicates that $\varepsilon > 1.10\%$ enters a trivial regime where the optimization task becomes too simple 
to exhibit meaningful scaling behavior.}
    \label{fig:advantage}
\end{figure}
\stepcounter{scounter}
\subsection{\thescounter: Possible quantum advantage for 3D spin glasses}
In the main text, we outlined a possible route toward achieving a quantum scaling advantage in future quantum annealers. 
This appears feasible for carefully chosen problem classes aligned with the strengths of quantum annealing, 
such as 3D spin-glass problems~\cite{King2023,Chowdhury2025}. 
Here, we present additional results motivating our choice of optimality thresholds 
and further comparing the scaling behavior of D-Wave and SBM, cf. Fig.~\ref{fig:advantage}.
Because D-Wave's topology does not natively support cubic lattices, we used an analytical minor embedding 
producing graphs with uniform chain length $2$. For SBM computations, the original instances were used directly.

\stepcounter{scounter}
\subsection{\thescounter: Classical and Quantum Hardware used}
All classical simulations were performed on a workstation equipped with dual Intel(R) Xeon(R)
Platinum 8462Y+ CPUs, 4 NVIDIA H100 SXM GPUs, and 1 TB of RAM.
All quantum experiments were performed on the D-Wave Advantage 4.1 QPU, located in the ``na-west-1''
region and accessed through the Leap cloud service. The physical properties of this QPU are summarized in Table~\ref{tab:advantage_parameters}.

\begin{table}[htbp]
    \centering
    \caption{D-Wave Advantage 4.1 parameters~\cite{adv4.1-properties}}
    \begin{tabular}{l l}
        \hline
        \textbf{Property}                                                                                       & \textbf{Value}                     \\
        \hline
        Model                                                                                                   & Advantage, performance update      \\
        Graph size                                                                                              & P16                                \\
        Number of qubits                                                                                        & 5627                               \\
        Number of couplers                                                                                      & 40279                              \\
        Qubit temperature                                                                                       & $15.4 \pm 0.1~\text{mK}$           \\
        $M_{\mathrm{AFM}}$ : Maximum mutual inductance for qubit pairs                                          & $1.647~\text{pH}$                  \\
        Quantum critical point for 1D chains                                                                    & $1.391~\text{GHz}$                 \\
        $L_q$ : Qubit inductance                                                                                & $372~\text{pH}$                    \\
        $C_q$ : Qubit capacitance                                                                               & $119~\text{fF}$                    \\
        $I_c$ : Qubit critical current                                                                          & $2.1~\mu\text{A}$                  \\
        Average single-qubit temperature                                                                        & $0.198$                            \\
        Ferromagnetic-problem freezeout                                                                         & $0.064$                            \\
        Single-qubit freezeout                                                                                  & $0.612$                            \\
        $\Phi_{\mathrm{CCJJ}}^{i}$ : Initial (at $s=0$) external flux on compound Josephson junctions\quad\quad & $-0.621~\Phi_0$                    \\
        $\Phi_{\mathrm{CCJJ}}^{f}$ : Final (at $s=1$) external flux on compound Josephson junctions \quad\quad  & $-0.717~\Phi_0$                    \\
        Readout time range                                                                                      & $17.0~\text{to}~235.0~\mu\text{s}$ \\
        Programming time                                                                                        & $\sim 14100~\mu\text{s}$           \\
        QPU-delay-time per sample                                                                               & $20.5~\mu\text{s}$                 \\
        Readout error rate                                                                                      & $\leq 0.001$                       \\
        \hline
    \end{tabular}
    \label{tab:advantage_parameters}
\end{table}

\end{document}